\documentclass[runningheads]{llncs}

\usepackage{graphicx}%
\usepackage{multirow}%
\usepackage{amsmath,amssymb,amsfonts}%
\usepackage{mathrsfs}%
\usepackage[title]{appendix}%
\usepackage{xcolor}%
\usepackage{textcomp}%
\usepackage{manyfoot}%
\usepackage{booktabs}%
\usepackage{subcaption}
\usepackage{algorithm}%
\usepackage{algorithmicx}%
\usepackage{algpseudocode}%
\usepackage{listings}%
\usepackage{svg}
\usepackage{tabularray}
\usepackage{tabularx} 
\usepackage{multirow} 
\usepackage{ragged2e} 
\usepackage{caption}
\usepackage[dvipsnames]{xcolor}
\usepackage{url}
\usepackage{hyperref}

\begin{document}

\title{A Data Annotation Requirements Representation and Specification (DARS)\vspace{-1em}}

\author{Yi Peng\inst{1}
\and
Hina Saeeda\inst{1}
\and
Hans-Martin Heyn\inst{1}
\and
\\
Jennifer Horkoff\inst{1}
\and
Eric Knauss\inst{1}
\and
Fredrik Warg\inst{2} 
}
\authorrunning{Y. Peng et al.}
\institute{The University of Gothenburg and Chalmers University of Technology, Gothenburg, Sweden, 
\email{\{yi.peng, jennifer.horkoff,eric.knauss\}@gu.se}, \email{\{hinasa,heyn\}@chalmers.se} \and
RISE, Borås, Sweden,
\email{fredrik.warg@ri.se}}

\maketitle 
\begin{abstract}
\vspace{-2em}
With the rise of AI-enabled cyber-physical systems, data annotation has become a critical yet often overlooked process in the development of these intelligent information systems. 
Existing work in requirements engineering (RE) has explored how requirements for AI systems and their data can be represented. 
However, related interviews with industry professionals show that data annotations and their related requirements introduce distinct challenges,  indicating a need for annotation-specific requirement representations. 
We propose the Data Annotation Requirements
Representation and Specification (DARS), including an Annotation Negotiation Card to align stakeholders on objectives and constraints, and a Scenario-Based Annotation Specification to express atomic and verifiable data annotation requirements. 
We evaluate DARS with an automotive perception case related to an ongoing project, and a mapping against 18 real-world data annotation error types. 
The results suggest that DARS mitigates root causes of completeness, accuracy, and consistency annotation errors. 
By integrating DARS into RE, this work improves the reliability of safety-critical systems using data annotations and demonstrates how engineering frameworks must evolve for data-dependent components of today's intelligent information systems.

\end{abstract}

\vspace{-0.5em}
\section{Introduction}\label{sec1}
The rise of pervasive AI technologies calls for a re-evaluation of information systems engineering practices for AI-enabled cyber-physical systems. 
Such systems increasingly include intelligent components, from locally trained machine learning (ML) models, to open-weight models and cloud-based foundation models~\cite{martinez2022software}. 
Data has become a critical, first-class system element, alongside code, databases and interfaces~\cite{habibullah2024scoping,sculley2015hidden}. 
This is particularly evident in safety-critical domains such as automated driving systems, where perception modules must interpret complex environments: a process with perceptual risks and uncertainties that traditional engineering methods struggle to mitigate~\cite{Li2023}. 
These perception modules rely on annotated data for training, for example images annotated with objects such as pedestrians or signage. 
In requirements engineering (RE) for systems with ML components, the field has identified significant challenges stemming from the complexity and non-determinism of AI~\cite{ahmad2023requirements,habibullah2024requirements,habiba2024mature,Chen2024}. 
Especially work on how to specify data sources, constraints, desired qualities, coverage, completeness~\cite{dey2023multi} and content requirements over data~\cite{peng2025mlreqs} recognizes that RE for AI systems must handle distinct data concerns.
However, specifying the data is only the first step; the actual preparation of that data, including their annotation, is often the most time-consuming phase of the lifecycle, requiring clarity and control to ensure reliable outcomes~\cite{Pucci2024}.
It is here that a critical gap remains: while general data requirements address properties of the dataset (e.g., size, balance, completeness), they often overlook the specific properties of the annotation. 
Our previous work shows that data annotation introduces unique challenges distinct from typical data requirements analysis, such as edge case annotation, unclear class definitions, and unclear purposes~\cite{hinaICSEJSS}, with recurring errors including attribute omission, or bounding-box errors~\cite{hinaCAIN}. 
We argue that many of these issues can be mitigated through annotation-specific requirements representation and negotiation.

In this work, we extend the argument for data requirements by establishing the need for dedicated data annotation requirements, including methods and structures to support their capture.
Beyond dataset properties, such as coverage, size, and completeness, annotation requirements must specify what to annotate (classes, attributes), how to handle difficult cases (e.g., pedestrian vs. a reflection of a pedestrian), expected accuracy and completeness of the annotation, tooling, format, and relevant ethical and legal constraints. 
We argue that this specific type of requirement demands representations distinct from existing requirements for AI-enabled systems and even from general data requirements. 
Although annotation requirements fall within data requirements, they call for unique considerations which we explore and exemplify in this paper.
Specifically, we introduce the \textit{Data Annotation Requirements Representation and Specification} (DARS), including an Annotation Negotiation Card and an associated Scenario-Based Annotation Specification template. 
Our work is inspired by existing approaches to negotiate~\cite{mlte_negotiation_card} and specify~\cite{peng2025mlreqs,dey2023multi} requirements for ML and data, but these approaches do not directly address challenges or errors related to data annotation requirements~\cite{hinaICSEJSS,hinaCAIN}.\par


\vspace{-1em}
\section{Background and Related Work}\label{sec:background}
We build on existing RE approaches for ML-enabled systems to develop a requirements solution for data annotation. 
Our approach synthesizes concepts from Machine Learning Test and Evaluation (MLTE) \cite{maffey2023mlteing}, the Multi-layered Data Requirements Framework~\cite{dey2023multi}, and our analysis of three established RE languages (EARS, Rupp's template, and Volere) for capturing ML system requirements~\cite{peng2025mlreqs}. We also provide an overview of related work in RE for AI and data.
None of these approaches are tailored specifically for data annotation, which is the gap we intend to address in this paper.

\noindent\textbf{MLTE and Quality Attribute Scenarios.} 
MLTE is a process and tool-set that supports interdisciplinary ML system development teams in testing and evaluating ML systems. 
The MLTE process is built around two key artifacts. 
The first is the Negotiation Card\footnote{Available at \url{https://mlte.readthedocs.io/en/latest/negotiation_card/}}~\cite{mlte_negotiation_card}, a high-level artifact that guides stakeholders in agreeing on mission and system-derived requirements that shape model development, including deployment environment, available data, and model requirements. 
It includes fields for system properties, data sources, and high-level data characteristics such as \emph{``labels and distribution''} 
and \emph{``labeling method''} (e.g., ``Hand labeled by a single domain expert''). 
The second artifact is the Quality Attribute (QA) scenario~\cite{brower2024using,bass2021software}, a methodology used by software architects and developers to guide design decisions and specify structural and behavioral requirements. 
QA scenarios support the negotiation by translating high-level goals into ML system requirements using six elements:
\vspace{-0.5em}
\begin{enumerate} \small
    \item \textbf{Stimulus:} A condition arriving at the system (e.g., an event).
    \item \textbf{Source of stimulus:} Where the stimulus comes from (e.g., external user)
    \item \textbf{Environment:} Set of circumstances in which the scenario takes place (e.g., normal operation).
    \item \textbf{Response:} Activity that occurs as the result of the arrival of the stimulus (e.g., deny access).
    \item \textbf{Artifact:} Target for the stimulus (e.g., data store).
    \item \textbf{Response measure:} Measures used to determine that the responses enumerated for the scenario have been achieved (e.g., latency).
\end{enumerate}
\noindent\textbf{The Multi-layered Data Requirements Framework.}
To ensure our solution for data annotation requirement is verifiable and traceable, we applied the multi-layered, evidence-driven framework of Dey and Lee~\cite{dey2023multi}. 
The framework conceptualizes a high-level problem space (capturing system goals and risks) projected onto a detailed data space (capturing data-specific characteristics). 
A key contribution is its data requirements template (Fig.~\ref{fig:multiLayeredTemplate}), which mandates `EVIDENCE' and `ACCEPTANCE CRITERIA' fields for each requirement, ensuring verifiability.
The template also enforces `VERTICAL TRACEABILITY' by linking each data requirement to a specific system-level goal in the Problem layer, which clarifies the rationale for the data.\par
\begin{figure}[b]
    \vspace{-0.4cm}
    \centering
    \includesvg[width=0.8\linewidth]{multi_layered_data_req_template.svg}
    \caption{Data requirements specification template from~\cite{dey2023multi}.}
    \label{fig:multiLayeredTemplate}
    \vspace{-1cm}
\end{figure}
\begin{figure}[t]
\vspace{-0.1cm}
    \centering
    \begin{subfigure}[b]{0.4\linewidth} 
        \centering 
        \includesvg[width=\linewidth]{EARS_template.svg}
        \caption*{(a)} 
        \label{fig:ears} 
    \end{subfigure}
    \hfill 
    \begin{subfigure}[b]{0.4\linewidth} 
        \centering 
        \resizebox{\linewidth}{4cm}{
        \includegraphics[width=\linewidth]{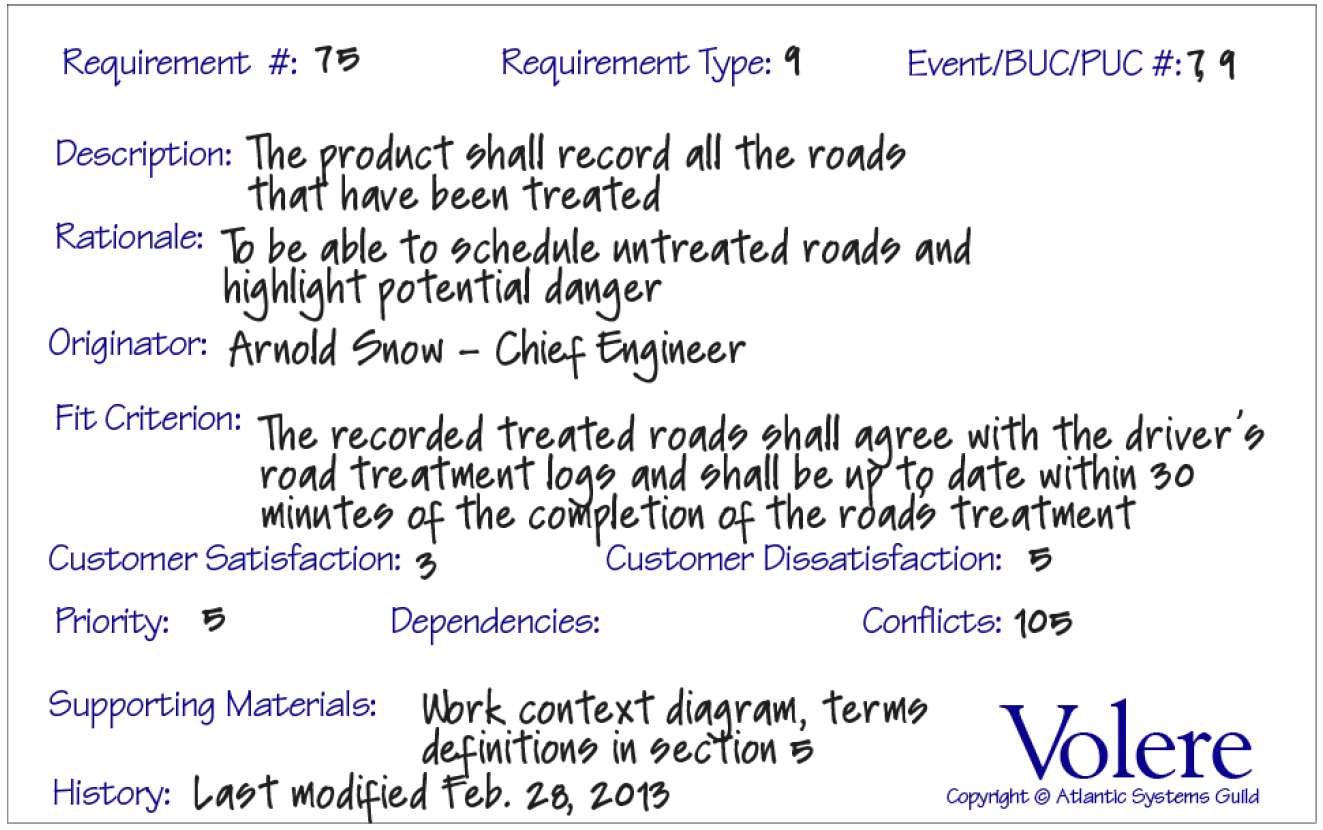}
    }
        \caption*{(b)} 
        \label{fig:rupp} 
    \end{subfigure}
\vspace{-0.5cm}
    \caption{(a) EARS~\cite{mavin2009ears}, 
    and (b) Volere~\cite{robertson1999volere}.}
    \label{fig:RE languages}
    \vspace{-0.5cm} 
\end{figure}

\noindent\textbf{Requirements Languages for ML Systems.} 
For specifying data annotation requirements, we took inspiration from established RE languages EARS~\cite{mavin2009ears}, Rupp's template~\cite{rupp2007requirements}, and Volere~\cite{robertson1999volere}.
Prior work~\cite{peng2025mlreqs} investigated their use for ML system requirements and found that the pattern-based and structured formats of EARS and Rupp's Template capture core technical details well. 
These included input-output specifications, training algorithms, integration interfaces, license constraints, intended users and use cases, and metrics for model evaluation. 
Volere proved more comprehensive, using its metadata atomic shell to capture critical ML context such as rational and fit criterion, along with external factors (e.g., assumptions, risks). 
Fig.~\ref{fig:RE languages} shows the core structures of EARS and Volere, as examples. 

\noindent\textbf{Challenges Specific to Data Annotation Requirements.}
Current research \cite{hinaICSEJSS} reveals that high-quality data annotation requirements are essential for reliable AI-enabled perception systems, yet industrial practice remains marked by ambiguity, inconsistencies, and resource limitations. 
These issues degrade annotation quality and system performance. 
The findings identify two dimensions for robust annotation requirements: \emph{representation requirements} that define what to capture, including precise classes and edge cases, and \emph{process requirements} that define how annotation is conducted, including guidelines for evolution and quality assurance. 
Current practice often fails on both fronts due to unclear definitions and reactive updates. 
Practitioners therefore recommend three key improvements: ensuring ethical compliance, refining guidelines iteratively, and embedding robust quality assurance such as multi-stage reviews and version control. 
Overall, the findings highlight the need to treat annotation requirements as evolving artifacts that explicitly define both content and process to ensure safety across the AI system development pipelines.

\noindent\textbf{Data Annotation Specific Errors.}
Recent work~\cite{hinaCAIN} systematizes data annotation challenges in AI-enabled perception systems into a taxonomy of 18 recurring errors. 
These errors fall into three quality dimensions: \emph{Completeness} (e.g., attribute omission, missing feedback loops), \emph{Accuracy} (e.g., wrong class labels, bounding-box errors), and \emph{Consistency} (e.g., inter-annotator disagreement, ambiguous instructions). 
The study suggests that these issues are systemic, driven by root causes such as vague guidelines, lack of domain knowledge, and misaligned hand-offs across the supply chain. 
Interviewed practitioners view this taxonomy as a ``failure-mode catalogue'' for diagnosing perception failures, highlighting the need for structured requirements to address these root causes~\cite{hinaCAIN}.

\noindent\textbf{Related Work from RE.}
Research in RE has produced a number of different ways to capture and represent requirements information, include structured user stories~\cite{lucassen2016use}, use case templates~\cite{cockburn1998basic}, and more formal approaches such as FRET~\cite{giannakopoulou2020formal}.
Although any structured approach may serve as inspiration and input to handle data annotation requirements, here we have focused on approaches which are specific to ML or data requirements, or which have been evaluated for this context, e.g., in~\cite{peng2025mlreqs}.  
\vspace{-1em}

\paragraph{RE for AI.} Requirements Engineering for Artificial Intelligence (AI) systems focuses on defining, managing, and validating the requirements needed to develop reliable, safe, and trustworthy data-driven software~\cite{habibullah2024requirements,ahmad2023requirements}. 
Unlike traditional systems, AI solutions rely on data, models, and probabilistic behavior rather than deterministic logic, which creates unique challenges for specifying functional and non-functional requirements~\cite{saeeda2025requirements}. 
RE for AI must address data quality, annotation guidelines, performance expectations, transparency, explainability, and continuous evolution of requirements as models are retrained~\cite{hinaICSEJSS}. 
It also involves aligning stakeholder needs with technical constraints, ethical considerations, regulatory demands and uncertainties of ML-based decision-making. 
Overall, RE for AI systems ensures that the system’s purpose, context, limitations, risks, and quality attributes are rigorously defined which then allows a dependable, safe, and fit for real-world use~\cite{hinaCAIN}. 
However, specific mechanisms for specifying data annotation requirements are still missing.

\vspace{-1em}

\paragraph{RE for Data.}
While RE for ML focuses on system-level behavior, a subset of research has begun to address eliciting, specifying, and verifying the data with dedicated methods, processes or tools~\cite{YiPeng2025DataChallengesAI}. 
Examples include knowledge-based approaches leverage ontologies~\cite{jiang2024dataset} or semantic webs~\cite{barzamini2022multi} to define domain concepts and evaluate dataset diversity~\cite{barzamini2022improving}. 
Tools like SEMKIS-DSL offer domain-specific languages for specifying dataset requirements~\cite{jahic2023semkis}. 
However, these approaches emphasize dataset collection and preparation, focusing on final quality properties rather than the annotation process. 
They overlook challenges of data annotation, such as inter-annotator disagreement, leaving a gap in specifying how data should be annotated to meet system goals.

\vspace{-0.5em}
\section{Data Annotation Requirements Representation and Specification (DARS)}\label{sec:result}
Based on the opportunities identified in our background analysis, we propose the Data Annotation Requirements Representation and Specification (DARS). Our analysis of frameworks like MLTE and the Multi-layered Data Requirements Framework revealed the need for at least two distinct artifacts: a high-level artifact for strategic alignment and a low-level artifact for a precise specification that is verifiable. 

The intended process, inspired by MLTE, is that a multidisciplinary team uses the high-level artifact to discuss and agree upon the goals and constraints of the data annotation process. This negotiation can be bi-directional: data providers define system needs, while annotation providers clarify operational constraints. The agreements recorded in this artifact then serve as the direct input for creating the low-level specification artifact, translating those goals to data annotation requirements. In our automotive perception example, the high-level artifact would capture a negotiation between different tiers, particularly where the party responsible for the perception component negotiates with the annotation provider about the specifics of the annotations for a particular dataset, including the underlying system and model objectives the dataset will likely fulfill.

DARS therefore is made up of two parts:
    a) The Annotation Negotiation Card (Card, for short): A document to align stakeholders on strategic goals, scope, and process before annotation begins.
   b) The Scenario-Based Annotation Requirement Specification (Specification, for short): A template for specifying atomic, verifiable annotation requirements based on scenarios derived from data quality attributes.

\subsection{The Annotation Negotiation Card}
The Annotation Negotiation Card is a high-level artifact that functions as a boundary object~\cite{kasauli2020charting} for all stakeholders (e.g., system owners, domain experts, annotation providers). Its purpose is to facilitate the capture of critical decisions that must be made before creating fine-grained annotation requirements. 

The structure is inspired by the MLTE Negotiation Card, which aligns interdisciplinary teams on ML model qualities and has been adapted for data annotation. Conceptually, this MLTE Negotiation Card extends the mapping that connects the ``Problem space'' to a general ``Data space'' from the Multi-layered data requirement framework. While the original framework's ``Data space'' leads to the specification of several requirement types (e.g., collection, representativeness, preparation, and annotation), our Annotation Negotiation Card refines this by linking the ``Problem space'' to a more specific ``Data Annotation space'' within the ``Data Space'', due to the unique set of unique challenges that necessitate a dedicated alignment artifact~\cite{hinaICSEJSS}.

This Card is designed to be filled out at the dataset level, capturing the specific agreements for a batch of data being handed over to annotators. While high-level system objectives may remain stable across multiple cards, specific operational context constraints or representation plans may vary per dataset. The fields on our card are derived from the business level concepts identified in our background work, such as the ``Problem Layer'' from the Multi-layered Framework and the high-level recommendations from practice. The card consists of six fields as shown in Table~\ref{tab:negotiationCard}.

{
\scriptsize
\begin{longtblr}[
  caption = {Annotation Negotiation Card},
  label = {tab:negotiationCard},
]{
  width = \linewidth,
  colspec = {Q[115]Q[390]Q[527]},
  vline{2-3} = {-}{},
  hline{1-2,8} = {-}{},
}
Field                                           & Description                                                                                                                                    & Inspiration                                                                                                                                                                                                                                                                                      \\
\textbf{1. System \& Model Objectives}             & A clear statement of the downstream system goals and the specific objectives of the ML model. It provides the context for why the data is needed and establishes how annotation quality impacts overall system performance.               & This field implements the ``Vertical Traceability'' concept from the Multi-layered Framework's ``Problem Layer'', linking the data annotation to a system-level goal.                                                                                                                                \\
\textbf{2. Operational Context \& Representation Plan}             & Specifies the valid operational boundaries (e.g., physical environment, user characteristics) of the system and a sampling plan to ensure annotation diversity and fair representation.          & Defining the context is a core concept from the Multi-layered Framework's ``Problem Layer''. The MLTE negotiation card also reports the distribution of data labels.                                                                                                                                       \\
\textbf{3. Ethical \& Legal Constraints}           & Explicit rules for handling PII (Personally Identifiable Information), anonymization, and regulatory compliance (e.g., GDPR).                  & This field is inspired by the MLTE Negotiation Card's sections on data rights and data policies and the recommendations of~\cite{hinaICSEJSS} to emphasize legal/ethical compliance.                                                                                                                     \\
\textbf{4. Tooling, Taxonomy, \& Hand-off}         & Specifies the exact technical standards: the annotation tools, the shared data format/ontology (e.g., ASAM OpenLABEL) including the full list of classes and attributes, and sensor calibration. & This field adapts the MLTE Negotiation Card's concerns about model deployment, including deployment platform and input output specification. It also draws upon the ideas of~\cite{hinaICSEJSS} on managing consistency between annotators.                                                  \\
\textbf{5. Quality Assurance \& Feedback Protocol} & Defines the process for quality assurance, such as review cadence (e.g., ``Weekly 10\% audit'') and a channel for closing the feedback loop.     & This is directly inspired by the ``Evidence Layer'' of the Multi-layered Framework. Recommendations from~\cite{hinaICSEJSS} also advocate for structured quality assurance mechanisms like multi-stage reviews, inter-annotator agreement checks, and feedback loops to identify and resolve ambiguity early. \\
\textbf{6. Guideline Governance}                   & Defines who owns the data annotation document, how it is versioned, and how changes are distributed.                                           & Volere's atomic requirement shell provides similar metadata on history and originator. Transparency and traceability in annotation workflows is another point from recommended practices in~\cite{hinaICSEJSS}.                                                                                                    
\end{longtblr}
}

\subsection{The Scenario-based Annotation Specification}
The Scenario-Based Annotation Specification is the template for defining individual annotation requirements. It is designed to be atomic, unambiguous, and verifiable. This structure translates abstract goals (from the Annotation Negotiation Card) into precise data annotation requirements for annotators that can be tested. 

We selected three scenario types based on the different requirement patterns of EARS and Rupp's template to cover the main categories of annotation requirements. Standard scenarios (inspired by EARS Ubiquitous pattern) define the main common-case rules. Edge-Case scenarios (inspired by EARS State-driven or Event-driven patterns) define the correct behavior for complex, rare, or ambiguous situations. Finally, Exception scenarios (inspired by the EARS Unwanted behavior pattern) explicitly define annotator actions under undesired or exclusionary conditions, e.g., outside the operational design domain (ODD) in the automotive case~\cite{ISO34503-2023}. The template (shown in Table~\ref{tab:annotationSpec}) synthesizes the key concepts of data annotation requirements from our background analysis.
{
\begin{table}[t]
\centering
\scriptsize
\caption{Scenario-Based Annotation Specification Template}
\label{tab:annotationSpec}
\renewcommand{\arraystretch}{1.3}
\begin{tabularx}{\linewidth}{
    >{\RaggedRight\bfseries}p{0.12\linewidth} 
    >{\RaggedRight}X 
    >{\RaggedRight\arraybackslash}X
} 
    \toprule
    Field & \textbf{Description} & \textbf{Inspiration} \\ 
    \midrule
    
    Req ID & 
    A unique identifier. & 
    General RE. How the identifier is defined can be based on a data annotation quality attribute (e.g., A-1 for~Accuracy-1). \\
    
    Scenario Type & 
    Standard, Edge-case, Exception & 
    Multi pattern-based approach of EARS and Rupp's template to capture the complexity and ambiguity of annotating data and avoid overlong, contradictory requirements that annotators cannot follow. \\
    
    Vertical trace & 
    A link back to a field in the Annotation Negotiation Card (e.g., ``Links to: System Goal 1'') & 
    This implements the Multi-layered Framework's concept of ``Vertical Traceability,'' ensuring every annotation requirement has a clear purpose. \\ 
    \midrule

    Stimulus & 
    The ``when''. Specific trigger for the requirement. (e.g., a pedestrian is 50\% occluded) & 
    \multirow{3}{=}{QA scenario. This also uses the structured-language format of \textbf{EARS} and \textbf{Rupp's Template}. The annotation requirement description is made up of: When \textless{}Stimulus\textgreater{}, in \textless{}Context\textgreater{}, the annotator shall \textless{}Action\textgreater{},~ with \textless{}Constraint\textgreater{}.} \\
    
    Context & 
    The ``given environment''. The environment where the stimulus occurs (e.g., urban environment, daytime etc) & 
    \\ 
    
    Response & 
    The ``then''. The precise action the annotator must take. Made up of an \textbf{Action} and a \textbf{Constraint (optional)}. & 
    \\ 
    \midrule

    Rationale & 
    An explanation of \textit{why} the rule exists (e.g., ``Tracker needs extended estimated box of occluded pedestrian for path prediction''). & 
    This incorporates the Volere template's emphasis on capturing Rationale to improve annotator understanding. \\
    
    {Acceptance criteria} & 
    The test for the requirement to show requirement fulfilled (e.g., ``Intersection over union 0.95 vs. ground truth''). & 
    This is the ``evidence'' component, fusing the QA Scenario's Response Measure with the Multi-layered Framework. \\
    
    Visual example & 
    Give a clear visual example. & 
    Recommendations from empirical results of~\cite{hinaICSEJSS}. \\ 
    \bottomrule
\end{tabularx}
\vspace{-1em}
\end{table}
}
\vspace{-1em}
\section{Evaluation}\label{sec:evaluation}
In alignment with the evaluation cycle of design science research~\cite{wieringa2014design}, we present a preliminary evaluation leveraging access to experts from both OEM (Original Equipment Manufacturer) and Tier-1 organizations to build a reasoned case for the proposed artifact's utility in its intended environment. We first demonstrate its application on a real-world Automatic Emergency Braking (AEB) system development scenario, using specific details from safety cases in~\cite{borg2023ergo}. We then validate the framework's problem-solving capacity by systematically mapping its components to the taxonomy of 18 empirically identified data annotation errors~\cite{hinaCAIN}, examining how the Card and Specification address the underlying root causes of these failures.
\vspace{-1em}
\subsection{Demonstration}
\vspace{-0.5em}
The example involves an OEM client developing a safety-critical perception system for urban driving and a Tier-1 supplier (acting as the data team/provider) collaborating with an external Annotation Vendor. The main challenge is distinguishing between road users (RUs) \textbf{Pedestrians} and \textbf{Riders} on micromobility vehicles like e-scooters and bikes. The project faces the following constraints: the OEM wants high recall for safety (no collisions), the data team needs specific class definitions for trajectory prediction (e.g., scooters move faster than walkers, so a vehicle needs to brake earlier), and the annotation vendor struggles with complex urban data (e.g., reflections, occlusions).\par
\noindent\textbf{Annotation Negotiation Card.}
To address the project challenges, the stakeholders utilize the annotation negotiation card to define boundaries before annotation begins. The filled-out card below establishes the necessary context to contribute to resolving the ambiguity between different types of road users for this specific dataset. Through this process, the card captures bi-directional agreements: the data provider defines the system goals and safety criticality, while the annotator defines the necessary data characteristics (e.g., privacy, balance) required to meet those goals. \par
\vspace{0.1em}
{
\scriptsize
\noindent \textbf{1. System \& Model Objectives}
\begin{itemize}
    \item \textit{System objective:} Enable safety-critical AEB in dense urban environments with high clutter artifacts (e.g., reflections and occlusions).
    \item \textit{Model objective:} Detect and classify road users for trajectory prediction and collision avoidance.
    \item \textit{Critical objective:} High recall. Missed detections of any RUs \textless{} 50m are safety failures.
    \item \textit{Downstream impact:} Class distinction (pedestrian vs. rider) is vital for velocity priors in the prediction model.
\end{itemize}

\noindent \textbf{2. Operation Context \& Representation Plan}
\begin{itemize}
    \item \textit{Operational context:} Dense urban centers with high object density.
    \item \textit{Representation plan:} To ensure high recall for all RUs and prevent selection bias, the dataset presented to the annotators must represent the following often missed data: road users (min. 10\% wheelchair users, min. 15\% e-scooter users, min. 5\% children), urban features (min. 10\% with glass storefronts), occlusion (min 5\% of RUs must be heavily occluded (\textgreater{} 50\%) to support amodal perception training)
\end{itemize}

\noindent \textbf{3. Ethical \& Legal Constraints}
\begin{itemize}
    \item \textit{PII anonymization:} 
    \begin{itemize}
        \item All human faces (pedestrians, riders, and visible drivers) and license plates must be anonymized (Gaussian blur) before the data is made available to annotators.
        \item If annotators encounter sensitive scenarios (e.g., accidents, personally identifiable private property), they must flag the frame for review and exclude it from dataset.
    \end{itemize}
    \item \textit{Regulatory compliance:} All data processing must occur on certified servers.
\end{itemize}

\noindent \textbf{4. Tooling, Taxonomy \& Hand-off}
\begin{itemize}
    \item \textit{Annotation tool:} Must support 2D bounding boxes with attribute tagging and dedicated quality assurance workflow.
    \item \textit{Sensor calibration:} Annotations must rely on synchronized sensor data. Sensor data shall be timestamped and calibration files are provided to validate 3D to 2D projections, ensuring alignment between LiDAR points and camera frames.
    \item \textit{Taxonomy definition:} 
   \begin{itemize}
        \item \textbf{Classes:}
        \begin{itemize}
            \item \texttt{Pedestrian}: Any live person on foot (walking, running, standing). Includes people pushing micromobility vehicles.
            \item \texttt{Rider}: Any person on a micromobility vehicle (bike, e-scooter, skateboard).
        \end{itemize}
        \item \textbf{Attributes:}
        \begin{itemize}
            \item \texttt{wheelchair}: Applicable to Rider class. Extra label if the person is using a wheelchair (for velocity prediction exception handling).
        \end{itemize}
    \end{itemize}
    \item \textit{Hand-off protocol:} Delivery in ASAM OpenLABEL format. Weekly batches of 1000 frames.
\end{itemize}

\noindent \textbf{5. Quality Assurance \& Feedback Protocol}
\begin{itemize}
    \item \textit{Acceptance thresholds:}
    \begin{itemize}
        \item Critical objective: $>$99\% recall for RUs (missed detections are grounds for batch rejection).
        \item Classification objective: $>$95\% precision for pedestrian vs. rider distinction.
    \end{itemize}
    \item \textit{Review cadence:} Weekly audit of 10\% of delivered assets by the quality assurance team.
    \item \textit{Feedback protocol:} 
    \begin{itemize}
        \item Edge cases \& exceptions: annotators must flag ambiguous frames.
        \item Calibration: weekly video calls between engineers in the data team and annotation vendor lead to review flagged cases.
    \end{itemize}
\end{itemize}

\noindent \textbf{6. Guideline Governance}
\begin{itemize}
    \item \textit{Ownership:} The data team lead is the owner of this annotation negotiation card and annotation requirement specifications. No changes are permitted without their direct approval.
    \item \textit{Versioning:} Guidelines shall follow semantic versioning (e.g., v1.0, v1.1).
    \begin{itemize}
        \item Every annotated batch must include metadata linking it to the specific Annotation Negotiation Card version and requirement specification version.
        \item Major version updates (e.g., v2.0) trigger mandatory re-training for all active annotators.
    \end{itemize}
    \item \textit{Distribution:} Updates to this annotation negotiation card and the set of annotation requirement specifications shall be pushed via a central notification system.
\end{itemize}
}
\noindent\textbf{Annotation Requirements Specification.}
Once the high-level objectives, constraints, and taxonomy are defined in the Annotation Negotiation Card, they must be translated into data annotation requirements for the annotation team. Our Specification template serves as this translation layer. Each requirement below is derived from a specific agreement in the Annotation Negotiation Card (e.g., taxonomy definitions, safety recall goals), and is structured to fulfill the specific constraints from the stakeholders.
We present examples of each scenario type in Fig.~\ref{fig:all_scenarios} to demonstrate how agreements from the Card can be translated into verifiable data annotation requirements.\par
\begin{figure}[h!]
    \centering
    \vspace{-1em}
    \begin{subfigure}{\linewidth}
        \centering
        \includesvg[width=0.75\linewidth]{standard_scenario.svg}
        \caption{Standard scenario data annotation requirements.}
        \vspace{-7.5em}
        \label{fig:standardScenario}
    \end{subfigure}
    
    \begin{subfigure}{\linewidth}
        \centering
        \includesvg[width=0.75\linewidth]{edge_case.svg}
        \caption{Edge-case scenario data annotation requirements.}
        \label{fig:edgeScenario}
    \end{subfigure}
    \vspace{0.5em}
    \begin{subfigure}{\linewidth}
        \centering
        \includesvg[width=\linewidth]{exception_scenario.svg}
        \caption{Exception scenario data annotation requirements.}
        \label{fig:exceptionScenario}
    \end{subfigure}
    \vspace{-2em}
    \caption{Overview of standard, edge-case, and exception scenario data annotation requirements with colors indicating \textcolor{Purple}{Stimulus}, \textcolor{ForestGreen}{Context}, \textcolor{MidnightBlue}{Action}, and \textcolor{orange}{Constraint}.}
    \vspace{-2em}
    \label{fig:all_scenarios}
\end{figure}
\vspace{-1.0em}
\paragraph{Standard scenarios:} Standard scenario annotation requirements define default scenarios featuring no exceptional or error conditions. In our case study, this includes requirements over the core pedestrian and rider classes to resolve any ambiguity. Two examples are presented in Fig.~\ref{fig:standardScenario}. A full production specification would include additional standard scenarios for other classes (e.g., gender/age group of the pedestrian), movement direction of the road user, and environmental conditions (e.g., lighting conditions).
\vspace{-1.0em}
\paragraph{Edge-case scenarios:} Edge-case scenario annotation requirements define rules for rare, atypical, or complex situations that fall at the boundary of expected behavior. Within the scope of our evaluation, we provide examples on reflections on glass storefronts and wheelchair users (a complex case where the classification ``rider'' conflicts with the expected speed estimation of ``slow'') in Fig.~\ref{fig:edgeScenario}. These are two representative examples, and many other edge cases exist in a production environment (e.g., unusual vehicle configurations).
\vspace{-1.0em}
\paragraph{Exception scenarios:} Exception scenario requirements define rules for handling data instances that should be excluded from the standard annotation process. Unlike edge-cases (which are rare but valid), exception scenarios can also involve triggers that invalidate the data for standard training purposes (e.g., quality failures, privacy risks, or out-of-ODD instances). Three exception scenario requirements are shown in Fig.~\ref{fig:exceptionScenario}.\par
\vspace{-1.5em}
\subsection{Systematic mitigation of annotation errors}
\vspace{-0.5em}
This section systematically evaluates the coverage of our solution against the taxonomy of 18 data annotation errors from Saeeda et al.~\cite{hinaCAIN} introduced in Sec.~\ref{sec:background}. Our evaluation maps these errors to specific components of the Card and Specification that contains the mechanism to prevent or mitigate it. We present this mapping in three parts, corresponding to the taxonomy's three main categories: annotation errors in completeness, accuracy and consistency. 

\textbf{Completeness errors} often stem from missing information (attributes, feedback) or missing processes (reviews, diversity checks). Our artifacts mainly mitigates these causes by forcing the definition of ``missing'' elements before data annotation begins. The Card addresses errors such like \textit{Selection bias} and \textit{Synchronization issues} by promoting discussion on explicit sampling quotas (\texttt{Operational context \& Representation plan}) and calibration protocols (\texttt{ Tooling, Taxonomy \& Hand-off}). The Specification tackles \textit{Attribute omission} by using \textless{}Stimulus\textgreater{} triggers that are linked to  mandatory \textless{}Response\textgreater{} attributes (Req ID: Exception-3, ``Stimulus: Wheelchair'' $\rightarrow$ ``Response: set wheelchair attribute label''), removing ambiguity. The \textit{Missing feedback loop} is addressed by the metadata \texttt{Acceptance criteria} in the Specification, ensuring a verification step that connects downstream data annotation activity informed by requirements to the \texttt{Quality Assurance \& Feedback Protocol} that is defined in the Card. We provide a detailed table (Table~\ref{tab:completenessMap}) that shows the link between the error and the solution for completeness errors.  Similar tables are available for Accuracy and Consistency errors in our online appendix~\footnote{\url{https://figshare.com/s/a024f7f69a8db80c54dc}, to be replaced with DOI.}.

{
\begin{table}[ht]
\centering
\scriptsize
\begin{talltblr}[
  caption = {Possible Mitigation of Data Annotation Completeness Errors (C: Annotation Negotiation Card, S: Scenario-Based Annotation Requirement Specification )},
  label   = {tab:completenessMap}
]{
  width = \linewidth,
  colspec = {X[0.8,l,t] X[1.6,l,t] X[2.8,l,t]},
  cell{2}{1} = {r=3}{}, 
  cell{5}{1} = {r=3}{}, 
  cell{5}{3} = {r=3}{}, 
  cell{8}{1} = {r=3}{}, 
  cell{8}{3} = {r=2}{}, 
  cell{11}{1} = {r=2}{}, 
  cell{13}{1} = {r=2}{}, 
  cell{13}{3} = {r=2}{}, 
  cell{15}{1} = {r=2}{}, 
  cell{15}{3} = {r=2}{},
  vline{2,3} = {-}{solid}, 
  hline{1,2,5,8,11,13,15,17} = {-}{solid},
  hline{3,4,6,7,9,10,12,14,16} = {2-3}{solid},
}
\textbf{Compl. Errors} & \textbf{Causes} & \textbf{Mitigation} \\
Attribute omission & Unclear schema & C: Tooling, taxonomy \& hand-off, forces definition of schema and list of wanted attributes (metadata). \\
 & Annotator fatigue & N/A \\
 & Weak validation/schema links & S: Links a stimulus (e.g., person is in wheelchair) to a response (e.g., set attribute label ``wheelchair''). \\
Missing feedback loop & No review/traceability & {C: Quality assurance \& Feedback protocol, mandates the review cadence and feedback channels. \\ S: ``Acceptance Criteria'' enforces checks.} \\
 & Weak communication & \\
 & Rare quality assurance reviews & \\
Privacy compliancy omissions & Absent anonymization & {C: Ethical \& legal constraints, set the privacy and legal rules. \\ S: apply exception scenario annotation requirement (e.g., Stimulus: ``Face is visible'', Response:``Apply blur''), use Vertical trace to link back to Ethical legal constraint.} \\
 & Unclear GDPR & \\
 & No audit logs & N/A\\
Unforeseen scenarios & No rare-case flagging/triage & S: Edge-case scenario type requirement is designed to capture these special events. \\
 & Incomplete guideline coverage & C: Operational context \& Representation plan, define the boundaries of in-scope and out-scope. \\
Selection bias & Over-sampling typical scenes & C: Operational context \& Representation plan explicitly requires sampling quotas to prevent this. \\
 & No diversity plan & \\
Synchr. issues & Unsynchoronized sensors & C: Tooling, taxonomy hand-off acts as agreed upon technical specs, requiring sensor sync definitions before data proc \\
 & Missing timestamps/calib metadata & 
\end{talltblr}
\vspace{-3em}
\end{table}
}

\textbf{Accuracy errors} are often caused by subjective judgment or vague instructions. Our solution mitigates these by providing objective requirements documented as actions that annotators can follow with precise constraints and enforcing a predefined quality assurance feedback process.

To address \textit{Wrong class label} and \textit{Bounding-box errors}, the Specification template provides measurable \texttt{Constraints} and uses \texttt{Exception scenarios} to explicitly identify anomalies in required annotation behavior, resolving boundary ambiguity. \textit{Granularity mismatch} can be prevented by the \texttt{Tooling, Taxonomy \& Hand-off} field from the Card, which mandates a shared ontology and depth definition. \textit{Bias-driven errors} are mitigated by the \texttt{Operational context \& Representation Plan}, which enforces sampling quotas and diversity targets of the dataset before data annotation occurs. The error \textit{Insufficient guidance}, caused by ``limited feedback'', is addressed by the \texttt{Quality Assurance \& Feedback Protocol} in the Card, which provides review loop to clarify any questions about the annotation requirements, while the \texttt{Guideline Governance} field ensures version control.\par

\textbf{Consistency errors} arise when annotators, teams or tools interpret data annotation differently. Our solution generally mitigates these issues by establishing a single source of truth that aligns interpretations and technical standards. 
The primary cause of \textit{Inter-annotator disagreement} is subjective interpretation and ambiguity. The Specification template could minimize this by breaking complex guidelines into atomic \texttt{Stimulus} to \texttt{Response} pairs. For example, to address the finding that ``different annotators will subjectively annotate the same object depending on distance or shape'' (e.g., one annotator labeling only the visible torso of an occluded pedestrian while another estimates the full body), the Specification defines the exact behavior: Stimulus: ``Pedestrian is partially occluded (e.g., lower half hidden)'', Context: ``Urban environment'', Action: ``Draw bounding box around full estimated extent of pedestrian'', Constraint: ``Box must include estimated position of feet on ground plane'', Rationale: ``Path prediction algorithms require full body estimation to calculate trajectory and ground contact point''. This replaces a subjective decision about ``how much to label'' with a deterministic rule derived from the system's need for path prediction. \textit{Ambiguous Instructions}, which are often caused by ``overlong or contradictory guidelines'' are resolved by the template's atomic structure. And with each annotation requirement, a \texttt{Vertical trace} back to sections in the annotation negotiation card is required, contradictory guidelines are exposed and resolved at the source, rather than confusing the annotator.

Errors like \textit{Lack of purpose knowledge} and \textit{Misaligned hand-offs} are communication failures. The Card addresses this by forcing agreements on \texttt{System \& Model Objectives}, ensuring every annotator understands the ``why'', and the \texttt{Hand-off Protocol} which acts as a contract for any transfer of annotation knowledge between teams or stakeholders. While there is a \textit{Lack of framework/ standards} that are internationally accepted, the Card instead enforces a project-dependent ``local standard'' in \texttt{Tooling, Taxonomy \& Hand-off Protocol}, where a shared taxonomy should be agreed upon during negotiation. This protocol also mitigates \textit{Cross-modality misalignment} by mandating that sensor synchronization and calibration rules (e.g., 3D-to-2D projection validation) are defined and agreed upon before annotation begins.\par

\textbf{Limitations:} Our artifacts cannot fully eliminate errors rooted in human or technical constraints. While possibly reducing cognitive load through structured specification, they do not prevent physical fatigue, over-trust, or subconscious bias. Similarly, they cannot fix infrastructure gaps like tool instability; however, the Card's \texttt{Tooling} and \texttt{Quality Assurance} protocols may mitigate this by optimizing tool selection and resource allocation. Lastly, privacy compliance relies ultimately on the tool provider's native capabilities.

\vspace{-1em}
\section{Discussion and Conclusion}\label{sec:disucssion}
\vspace{-0.5em}
Our evaluation shows that DARS provides a systematic mechanism to mitigate at least one root cause for every annotation error in the 18-item data annotation error taxonomy, indicating the potential to address completeness, accuracy and consistency errors in the industrial annotation process.
Our analysis further shows that the Card addresses 29 of the 44 identified root causes (approx. 66\%), whereas the Specification addresses 18 (approx. 40\%), suggesting that many errors stem from strategic and process misalignment rather than execution failures. 
Thus, data annotation is a process where agreement on the task is as critical as the quality of the actual labeling. 
While MLTE connects the ML system problem space to a general data space, our findings suggests the need for a dedicated data annotation space with dedicated artifacts. 
DARS offers this structure to resolve strategic misalignment through the Card before they propagate to the data level, with vertical traceability ensuring that the ``why'' in the Card is embedded in every atomic requirement via the Specification.
These principles may become increasingly relevant as industry shifts toward automated and AI-aided workflows (e.g.,~\cite{xie2024automating,sutharsan2023smart,tejani2022performance}). 
In such semi-automated pipelines, the human annotator shifts from creator to verifier, making the Specification a ground truth definition for auditing the automated output. 
The Card's strategic constraints may serve as guardrails to prevent biases from AI model used to generate annotations from propagating into the final dataset. 
Future iterations of DARS therefore could further include specific verification triggers for AI-generated labels.

\textbf{Threats to Validity.}
The primary threat is that our evaluation utilizes a data annotation error taxonomy and demonstration case derived exclusively from the automotive perception domain. While this sector is relatively advanced in annotation practices, specific challenges in other domains (e.g., ambiguity in medical imaging or subjectivity in natural language processing) may differ. We mitigate this by basing our artifact design on domain-agnostic RE frameworks and representations, but further validation in non-automotive contexts is required to claim broader applicability. Furthermore, while it was designed to reflect real-world constraints based on a safety case for a ML component in a pedestrian AEB~\cite{borg2023ergo}, the demonstration case is simplified from reality.

Our evaluation also relies on a mapping between identified errors and artifact features. This mapping is analytical and may reflect researcher bias regarding the potential efficacy of the solution versus its actual performance in an industrial environment. However, initial, informal feedback from our automotive project partners has been positive. This threat shall be further addressed in future work through more systematic expert feedback and deployment in industry.

\textbf{Future Work.}
Future work will focus on industrial validation beginning with expert workshops and focus groups with automotive industry partners to refine the artifacts. We then aim to deploy DARS in active annotation pipelines to measure its impact on the data annotation process. Furthermore, inspired by MLTE implementation, we plan to develop a dedicated software tool-set that allows practitioners to create, edit and manage these artifacts. Finally, we will explore the generalizability of our findings to other domains beyond perception systems in the automotive industry.\par
\textbf{Conclusion.} 
Data annotation errors can cascade into critical failures in AI-enabled perception systems. 
We propose DARS consisting of an Annotation Negotiation Card and a Scenario-based Annotation Specification template. 
Evaluated against a taxonomy of 18 real-world annotation errors, DARS shows potential to mitigate root causes of annotation completeness, accuracy and consistency errors. 
This work highlights that high-quality annotated data depends not only on labeling effort, but on explicit strategic alignment and verifiable, human-centric annotation requirements. 
Future work will validate these artifacts in industrial pipelines and explore their generalizability.
\vspace{-0.5cm}
\begin{credits}
\subsubsection{\ackname} Supported by the Swedish Research Council (VR) iNFoRM Project and the Wallenberg AI, Autonomous Systems and Software Program (WASP).
\end{credits}

\bibliographystyle{plainurl}
\bibliography{sn-bibliography}

\end{document}